\documentclass{IEEEtran}
\usepackage{cite}
\usepackage{amsmath,amssymb,amsfonts}
\usepackage{graphicx}
\usepackage{textcomp,nicefrac}
\usepackage{float}

\def\EP#1{\times10^{#1}}
\def\lumiunit{\mbox{~cm}^{-2}\mbox{s}^{-1}}
\def\fbinv{\mbox{~fb}^{-1}}

\begin{document}
\title{Performance of the Unified Readout System of Belle II}
\author{%
  Mikihiko Nakao,
  Ryosuke Itoh,
  Satoru Yamada,
  Soh Y. Suzuki,
  Tomoyuki Konno,
  Qi-Dong Zhou,
  Takuto Kunigo,
  Ryohei Sugiura,
  Seokhee Park,
  Zhen-An Liu,
  Jingzhou Zhao,
  Igor Konorov,
  Dmytro Levit,
  Katsuro Nakamura,
  Hikaru Tanigawa,
  Nanae Taniguchi,
  Tomohisa Uchida,
  Kurtis Nishimura,
  Oskar Hartbrich,
  Yun-Tsung Lai,
  Masayoshi Shoji,
  Alexander Kuzmin,
  Vladimir Zhulanov,
  Brandon Kunkler,
  Isar Mostafanezhad,
  Hideyuki Nakazawa,
  and
  Yuji Unno% <-this % to avoid a space
  \thanks{Manuscript submitted on October 29, 2020, revised on February
    25, 2021.}%
  \thanks{%
    M. Nakao (e-mail: mikihiko.nakao@kek.jp),
    R. Itoh, S.Y. Suzuki, K. Nakamura are
    with KEK, High Energy Accelerator Research
    Organization, and SOKENDAI,
    % School of High Energy Accelerator Science,
    the Graduate University for Advanced
    Studies, Tsukuba, 305-0801 Japan.}
  \thanks{%
    S. Yamada, T. Kunigo,
    S. Park, N. Taniguchi, T. Uchida, and M. Shoji
    are with KEK, High Energy Accelerator Research
    Organization, Tsukuba, 305-0801 Japan.}    
  \thanks{%
    T. Konno is with Kitasato University, Sagamihara, 252-0373
    Japan.}
  \thanks{%
    Q.D. Zhou is with Nagoya University, Nagoya, 464-8602 Japan.}
  \thanks{%
    R. Sugiura and H. Tanigawa are with Tokyo University, Tokyo,
    113-0033 Japan.}
  \thanks{%
    Z.-A. Liu and J. Zhao are with Institute of High Energy Physics,
    Chinese Academy of Sciences, Beijing, 100049 China}
  \thanks{%
    I. Konorov and D. Levit are with 
    Technische Universit\"at M\"unchen, Garching, 85748 Germany}
  \thanks{%
    K. Nishimura, O. Hartbrich and I. Mostafanezhad are with University
    of Hawaii, Honolulu, 96822 USA.}
  \thanks{%
    Y.-T. Lai is with Kavli Institute for the Physics and Mathematics of
    the Universe, University of Tokyo, Kashiwa, 277-8583 Japan.}
  \thanks{%
    A. Kuzmin and V. Zhulanov are with Budker Institute of Nuclear
    Physics SB RAS, Novosibirsk, 630090 Russia.}
  \thanks{%
    B. Kunkler is with Indiana University, Bloomington, 47408 USA.}
  \thanks{%
    H. Nakazawa is with National Taiwan University, Taipei, 10617 Taiwan.}
  \thanks{%
    Y. Unno is with Hanyang University, Seoul, 04763 Korea.}
}

\maketitle

\begin{abstract}

The Belle II experiment at the SuperKEKB collider at KEK, Tsukuba, Japan has
successfully started taking data with the full detector in March
2019.  Belle II is a luminosity frontier experiment of the new
generation to search for physics beyond the Standard Model of elementary
particles, from precision measurements of a huge number of
{\boldmath$B$} and charm mesons and tau leptons.  In order to read out
the events at a high rate from the seven subdetectors of Belle II, we
adopt a highly unified readout system, including a unified trigger
timing distribution system (TTD), a unified high speed data link system
(Belle2link), and a common backend system to receive Belle2link data.
Each subdetector frontend readout system has a field-programmable gate
array (FPGA) in which unified firmware components of the TTD receiver
and Belle2link transmitter are embedded.  The system is designed for
data taking at a trigger rate up to 30 kHz with a dead-time fraction of
about 1\% in the frontend readout system.  The trigger rate
%at the current accelerator commissioning stage
is still much lower than
our design.  However, the
background level is already high due to the initial vacuum condition and
other accelerator parameters, and it is the most limiting factor of the
accelerator and detector operation.  Hence the occupancy and radiation effects
to the frontend electronics are rather severe, and they cause various
kind of instabilities.  We present the performance of the system,
including the achieved trigger rate, dead-time fraction, stability, and
discuss the experience gained during the operation.

% (omit from abstract)
% We share the operation time between the accelerator commissioning and
% physics data taking.

\end{abstract}

\begin{IEEEkeywords}
  % key words or phrases in alphabetical order, separated by commas
  % list of suggested keywords from
  % http://www.ieee.org/organizations/pubs/ani_prod/keywrd98.txt
  % (i.e., taxonomy_v101.pdf)
  Availability,
  Centralized control,
  Data acquisition,
  Data transfer,
  High energy physics, % not in list
  Radiation effects,
  System integration,
  Synchronization  % not in list

\end{IEEEkeywords}

\section{Introduction}
\label{sec:intro}

Belle II experiment~\cite{tdr} at the SuperKEKB $e^+e^-$
collider~\cite{skb} at KEK, Tsukuba, Japan has successfully started the
data taking %to play the role of the new generation luminosity frontier
%experiment
to search for physics beyond the Standard Model of elementary
particles.  The goal of the Belle II experiment is to collect
an unprecedented 50 ab$^{-1}$ integrated luminosity, mostly at the
$\Upsilon(4S)$ resonance, to study and search for a wide range of $B$
meson decays, charm meson decays, $\tau$ lepton decays, and hypothetical
particles such as those expected from the dark sector.  The design
instantaneous luminosity is $8\times10^{35}$~cm$^{-2}$s$^{-1}$, 40
times higher than the highest luminosity achieved by its predecessor,
KEKB.  Thanks to the clean environment of the $e^+e^-$ collision, the
events are triggered with a single level (level-1) trigger system with
a trigger efficiency greater than 99\% for most of the $B$ meson decay
modes.  The level-1 trigger rate is designed to be up to 30 kHz, which
includes about 1 kHz each of $B$-meson-pair, charm-pair, and
$\tau$-lepton-pair events.

The Belle II detector consists of seven subdetectors: a pixel detector
(PXD) and a silicon-strip vertex detector (SVD) for vertex
reconstruction, a central drift chamber (CDC) for charged track
reconstruction, a time-of-propagation counter (TOP) and an aerogel
ring imaging Cherenkov
detector (ARICH) for charged hadron identification, an electromagnetic
calorimeter (ECL) for photon detection and electron identification, and
a $K_L$ and muon detector (KLM) in the return yoke of the 1.5 T solenoid
coil.  The event is read out upon each level-1 trigger decision based
mostly on CDC and ECL trigger information, which is given within a
latency of about 5 $\mu$s.  All the detector data are digitized inside
or nearby the detector, and collected by the data acquisition system.

The first physics run was in 2018 with the Belle II detector without the
vertex detectors, under the so-called phase 2 operation.  The main
purposes were the commissioning of the accelerator, evaluation of the
background condition for the vertex detectors, and initial physics
programs with a low-multiplicity trigger condition and with no
requirement on the precise vertex information.  The main physics
program, so-called phase 3, has successfully started in 2019 with
the full Belle II detector.  Although the luminosity is still far below
the design, it reached the peak of $2.4\EP{34}\lumiunit$,
already exceeding the previous record established by KEKB.  Belle II has
collected $74\fbinv$ of data, with an overall efficiency of about 84\%,
as discussed later.

In this paper, we first briefly describe the unified readout system of
Belle II, and then the performance of the system and various problems we
experienced in the first two years of operation.

\section{Unified Readout System}

In order to read out the events from the seven subdetectors, we adopt a
highly unified readout system~\cite{nakao}\cite{yamada}, including a
unified trigger timing distribution (TTD) system for the entire Belle II
detector, a unified high speed data link system called Belle2link,
which is used by all subdetectors except PXD, and a common backend
system called COPPER to receive the Belle2link data.  Every
subdetector frontend electronics (FEE) device has an FPGA in which the
unified firmware components of TTD receiver and Belle2link transmitter
are embedded.

The system aims for taking data at 30 kHz trigger rate with a dead-time
fraction of about 1\% from the frontend readout system.  The read-out
data are sent to the backend data acquisition system comprised of the
event builder, high level trigger and storage system.  The schematic
view of the Belle data acquisition system is given in Fig.~1.

\begin{figure*}[ht]
\centerline{%
  \resizebox{0.95\textwidth}{!}{%
    \includegraphics{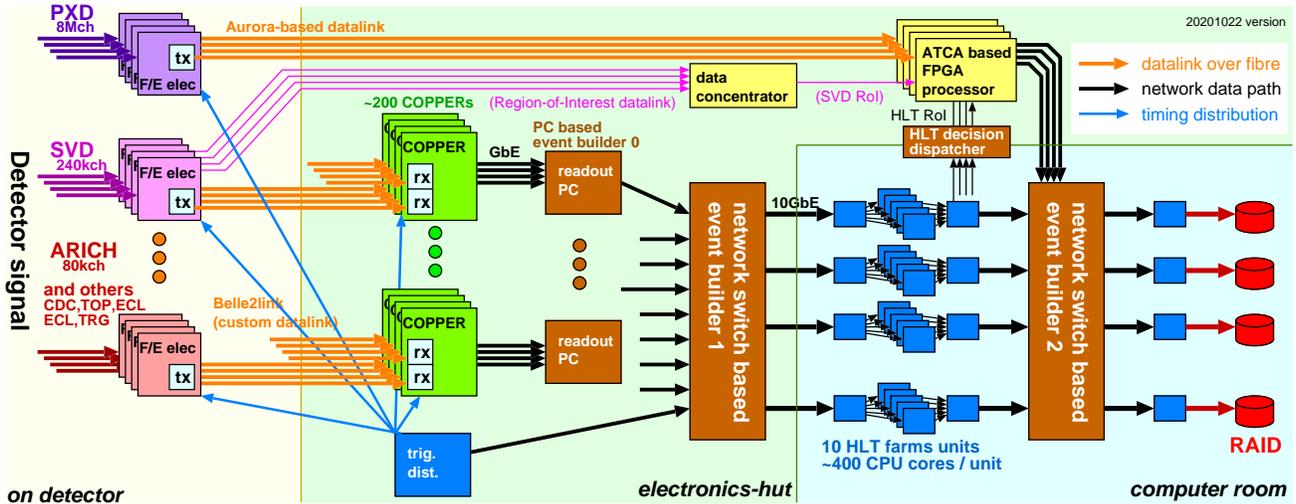}}}
\caption{Schematic view of the Belle II data acquisition system.}
\end{figure*}

\subsection{Unified Trigger Timing Distribution}

The TTD system~\cite{ttd} consists of a network of the frontend-timing-switch
(FTSW) modules in a tree structure, to distribute the system clock,
level-l trigger timing, and other information for fast control to a
large number of FEE boards and COPPER modules, and to collect their
status.
%
% The TTD system~\cite{ttd} is a tree-like connection of the frontend-timing-switch
% (FTSW) modules to a large number of FEE boards and COPPER modules, to
% distribute the system clock, level-l trigger timing and other
% information for fast control and to collect the status of FEE and
% COPPER.
%
The system clock of 127 MHz is generated from the
509 MHz radio frequency (RF) of SuperKEKB, and is directly distributed
using a dedicated line.  The remaining signals are embedded in a
bidirectional serial link of 254 Mbps using a custom protocol called
b2tt.  These signals are transmitted as low voltage differential
signaling (LVDS) signals over a low-cost category-7 network cable for
most of the short distance
connections, or over two pairs of multimode fibers for the long distance
connections
between the stations on the detector and the electronics-hut where the
center of the TTD system resides.

\begin{figure*}[ht]
  \centerline{%
    \resizebox{0.95\textwidth}{!}{%
      \includegraphics{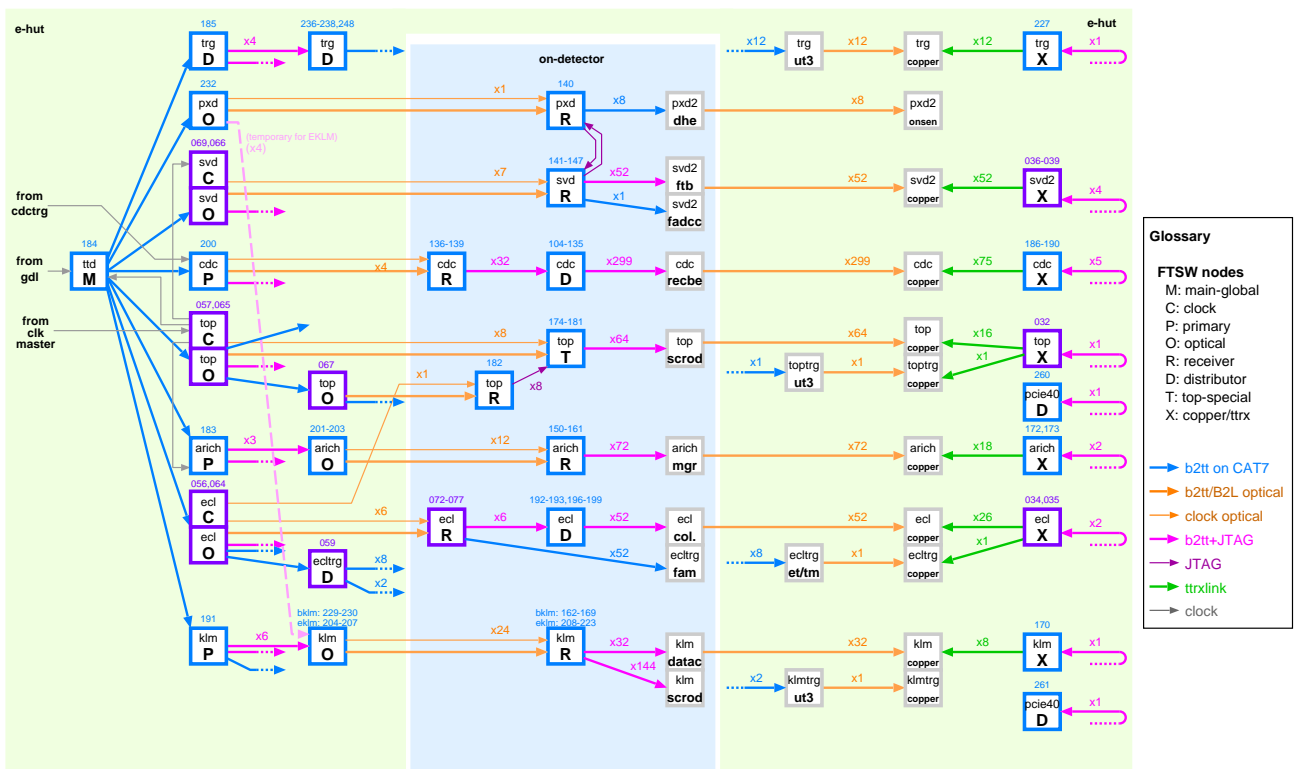}}}
  \caption{Schematic view of the trigger timing distribution tree.  The
    symbols M, C, O, P, R, D, T, X correspond to different firmware used
    by FTSW modules.}
  \label{fig:ttd}
\end{figure*}

The FTSW module~\cite{ttd} is a multi-purpose double-width 6U-height
VMEbus~\cite{vme} module equipped with a Xilinx~\cite{xilinx} Virtex-5
FPGA and 24 RJ-45 connectors.  Four of these connectors have dedicated
purposes: one for the connection to uplink, one for programming of the
FPGA of the FTSW using JTAG~\cite{jtag}, one for Ethernet (unused), and
one for multipurpose LVDS input or output; and remaining 20 connectors
are used for distribution.  The bottom 4 or 8 distribution RJ-45
connectors can be replaced with an FMC daughter card with 2- or 8-port
SFP optical transceivers, to receive or distribute the b2tt serial-link
signals.  Up to 4 stages of cascaded connections of FTSW modules are
used to deliver the TTD signal to more than 1,000 destinations of FEE
boards and other systems as shown in Fig.~\ref{fig:ttd}.

The FTSW module is also used to deliver the JTAG signals to the frontend
boards, first encoded in the b2tt protocol and delivered to the last
step of the FTSW tree, and then transmitted as LVDS level signals to the
FEE over another category-7 cable.  Therefore, a typical FTSW module on
the detector is receiving the b2tt serial link over 2 pairs of fibers
and connected with 8 FEE boards for timing distribution and JTAG
programming.

The TTD system distributes the level-1 trigger signal with the event
number, timestamp and trigger type.  The timestamp is a unique and
monotonically incremented 59-bit value for every event, and it is saved in the
data to be used later to detect event mismatch and data errors at
various stages of the readout chain.  The trigger type is used to
dynamically change the readout operation of FEE depending on the trigger
source.  The trigger interval is controlled by a programmable interval
counter and an emulation logic of the SVD FEE to avoid the overflow in
the SVD FEE, which has the most timing-critical condition among
subdetectors.
% In addition, busy signals are accepted to pose the back
% pressure from the backend data transport and some of the FEE systems.
In addition, busy signals are collected to temporarily pause the trigger
distribution as the back pressure from the backend data transport and
some of the FEE systems.

At the same time, the TTD tree is used to collect and summarize the
status of the readout system, including error information, number of
processed events, and status of the SEU mitigation (see section~\ref{sec:seu}).
Each connection can be masked or reset
remotely, to avoid spurious information from unused or malfunctioning
links.  In addition to the FEE, the TTD system also distributes various
fast timing information to subdetector and global trigger processors,
luminosity counters, and beam background monitors.

\subsection{Unified Data Link --- Belle2link}

The Belle2link is a bi-directional custom high-speed serial link
protocol to collect the data read out at the FEE~\cite{b2l}.  It uses
the 8b10b encoded
%GTP or GTX
high speed serial link function of the
Xilinx FPGA.  The raw bit rate is 2.54 Gbps, driven by the system clock,
but the payload bandwidth is limited to about 1 Gbps at the FEE, mainly
because the bandwidth is limited by the COPPER backend interface design
and because there is no
back pressure from the COPPER to the FEE for simplicity.

The receiver of the Belle2link is a single-channel optical receiver card
called HSLB~\cite{b2l} equipped with a Xilinx Virtex-5 FPGA. Up to 4 HSLB
cards are mounted on a COPPER module, which is a 9U-height VMEbus
board.  The COPPER module is driven by a processor card running Linux
operating system on an Intel x86 processor.  The COPPER module is a
multi-purpose platform, which is also used by other experiments with
different daughter cards instead of the HSLB.

The event fragment data sent from the FEE is checked for errors, and then
copied to the FIFO buffer of the COPPER.  The COPPER module then combines
the event fragments and makes a direct memory access (DMA) transfer to
the processor.  The processor is used to make a minimal formatting and
send the data to the next stage through a Gigabit Ethernet connection.

\begin{figure}[hb]
  \centerline{%
    \resizebox{0.4\textwidth}{!}{%
      \includegraphics{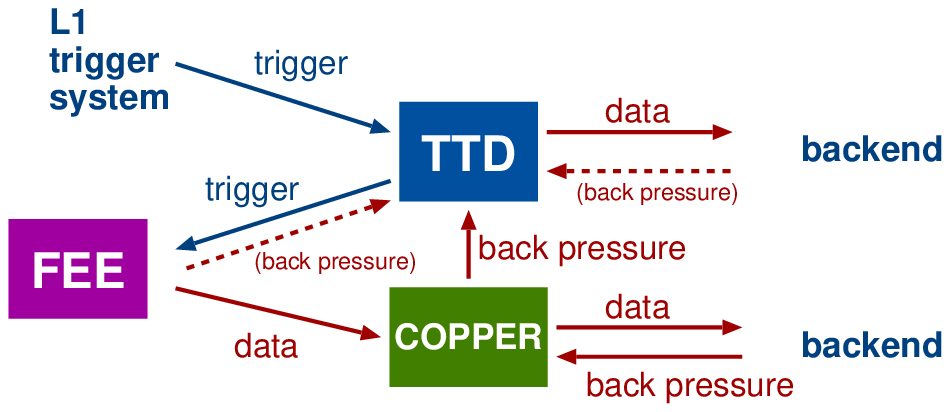}}}
  \caption{Trigger, data, and back-pressure paths of the unified readout
    system.}
  \label{fig:trg-dat-bsy}
\end{figure}

The COPPER board also is equipped with a trigger card for the connection
to the
TTD system.  A programmable threshold is set to the FIFO buffer, and
when the data exceeds the threshold, a back pressure is sent to the TTD
system.  The entire path of the trigger distribution, data collection,
and back pressure is illustrated in Fig.~\ref{fig:trg-dat-bsy}.

The HSLB card also serves as interface to the FEE for read and
write access to the 32-bit registers mapped onto a 16-bit address space.
These registers are used to configure individual boards, such as setting
the threshold or parameters that are needed for feature extraction from
the read-out signal waveform, and reading the individual status of the
FEE boards, such as the voltage of the supplied power or temperature of
the FPGA.

\subsection{Subdetector Frontend Electronics}

Although the TTD and Belle2link are common, the requirements and hence
the hardware designs for the FEE boards differ among subdetectors.  The
simplest example is the CDC FEE board, which performs 48-channel
preamplification, shaping and analog-to-digital conversion at a 31 MHz
sampling cycle on the board, and time-to-digital conversion with a
1$\,$ns least-significant-bit realized inside a Xilinx Virtex-5 FPGA.
Other subdetectors require additional preprocessing steps, typically
using an external analog-to-digital conversion circuit and a digital
logic built in another FPGA.  The most complex FEE is the one for
TOP~\cite{topfee}, which is built upon the Xilinx ZYNQ system-on-chip device with a
Xilinx 7-series FPGA core for timing-critical logic and an Arm processor
core for pipelined data processing.
%which is used as a
%part of the pipeline to process the data.
All subdetector frontend
electronics are based on one of the Xilinx FPGA devices (Virtex-5,
Spartan-6, Virtex-6, Kintex-7 or ZYNQ), with the exception of the
flash-ADC controller board of SVD with the Stratix IV FPGA of
Intel (Altera)~\cite{altera}.

\subsection{PXD and Backend}

The data read out by the COPPER are collected and built into a raw event
without the PXD data.  The raw events are fed into
the high level trigger (HLT) computing nodes, where the full event
reconstruction is made to filter the events by up to a factor of 5.  The
HLT consists of multiple streams, each of which is an independent
parallel processing block of computing nodes and dedicated input and
output nodes.  The
number of HLT streams has been and will be increased in a staged way;
HLT has been operated with 9 streams until summer 2020, and the stream
10 was added during summer shutdown.

% The data read out by the COPPER are collected and built into a raw event
% without the PXD data.  The raw events are fed into one of the streams of
% the high level trigger (HLT) computing nodes, where the full event
% reconstruction is made to filter the events by up to a factor of 5.  The
% number of HLT streams has been and will be increased in a staged way;
% HLT has been operated with 9 streams until summer 2020, and the stream
% 10 was added during summer shutdown.

The PXD data is not combined at this stage for two reasons.  First, the
data size, which is an order of magnitude larger than the sum of the
rest, is beyond the limited bandwidth of the COPPER based unified
readout system.  Second, PXD information does not contribute to the HLT event
filtering.  Contrary, we use the reconstructed charged tracks at HLT to
reduce the PXD data by an order of magnitude by only saving the
region-of-interest subset, and make the final event building before
saving the data into a storage device.

\section{Operation and Performance}

The phase 3 operation of Belle II has started in March 2019, which is
the final phase of the commissioning with all subdetectors and
accelerator components.  In 2019, it continued until July, and then
after a summer break, resumed from October till December.  The run in
2020 started in February and continued until July, and resumed in
October to end in December.

\subsection{Operating Conditions}

SuperKEKB and Belle II are nominally operated to continuously accumulate
physics data, except for the scheduled half-day accelerator maintenance
every 2 weeks.  However, the current priority is in improving the peak
luminosity
rather than maximizing the integrated luminosity.  Until summer 2020,
day time of weekdays were usually devoted to the accelerator studies, and
night time and weekend were used for physics data taking.

% When SuperKEKB and Belle II are operated, it continues for 24 hours of 7
% days per week, except for the scheduled half-day accelerator maintenance 
% every 2 weeks.  The current priority is in improving the peak luminosity
% rather than maximizing the integrated luminosity.  Until summer 2020,
% day time of weekdays are usually devoted to the accelerator studies, and
% night time and weekend are used for physics data taking.

In the current operation, the beam current is limited to
keep the beam background condition below the limit on the
integrated dose to the photon detector of TOP.  As a result, the
trigger rate is still far below the design.  Typical level-1 trigger
rate around the end of the latest run period was around 4 kHz, whereas
the
%expected trigger rate is 10 kHz for the full luminosity and the
design trigger rate of the system at the full luminosity is 30 kHz.

The time for accelerator studies are used to operate the data
acquisition system with 30kHz dummy random triggers with intervals of a
pseudo Poisson distribution.  Since high voltage is not
applied to the subdetectors, threshold is lowered for CDC to generate
data with a reasonable occupancy.  This dummy trigger operation has been
useful to keep updating the firmware and software to improve the
performance and stability.

\begin{figure}[ht]
  \centerline{%
    \resizebox{0.5\textwidth}{!}{%
      \includegraphics{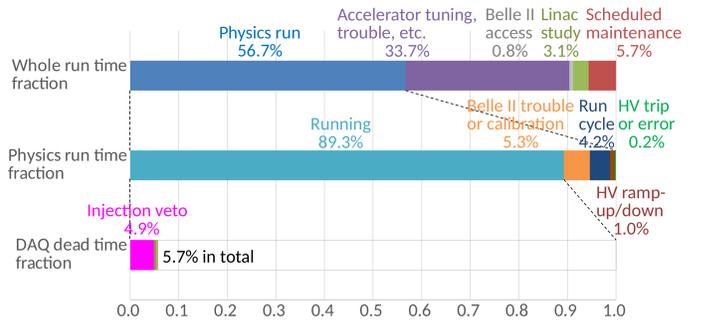}}}
  \caption{Overall data acquisition efficiency in the run
    period from February to July, 2020.}
  \label{fig:effsummary}
\end{figure}

A summary of the operation and dead-time fraction in 2020 is given in
Fig.~\ref{fig:effsummary}, with an overall efficiency of 84.2\%.

\subsection{Dead-time Fraction}

The largest dead-time fraction during the stable operation comes from
the veto window after the beam injection.  In the continuous injection
mode, the linac injects the beam to the main ring during the run to keep
the beam current and other accelerator conditions constant.  The beam
injection occurs at a rate of 25 Hz at most.  Each beam injection heavily disturbs the
succeeding stored bunches near the injected bunch, and it also disturbs the
stored bunches in the entire ring to some extent.  In order to reduce the spurious
triggers due to the disturbed bunches, the level-1 trigger is vetoed for
all the bunches for a short period right after the injection, and for
the bunches near the injection for a longer period.  The veto length is
tuned to minimize the dead-time fraction, which is about 5\%.

% The beam injection occurs at a cycle of 25 Hz at most.  The level-1
% trigger is entirely masked for a short period right after the injection
% timing, and then for the timing of the injected beam bunch for a longer
% period.  The veto length is tuned to avoid spurious triggers due to
% the injection background, and in total about 5\% of time is vetoed.

The second major dead time comes from the run restart cycle, which
typically takes about 2.5 minutes, but may take longer depending on the
situation.  We pose an 8-hour limit on the run length, but most of the
runs are stopped much earlier by the loss of the beam or by an error in
the data acquisition.

The dead-time fraction from the data acquisition system is less than
1\%.  Two dominant contributions come from the trigger throttle and
slow-down of the readout software somewhere in the chain, e.g.\ caused
by a flood of log messages when partially broken data are detected.  Otherwise the
dead-time fraction due to the data acquisition system itself is
negligibly small.

\begin{figure*}[ht]
  \centerline{%
    \resizebox{0.95\textwidth}{!}{%
      \includegraphics{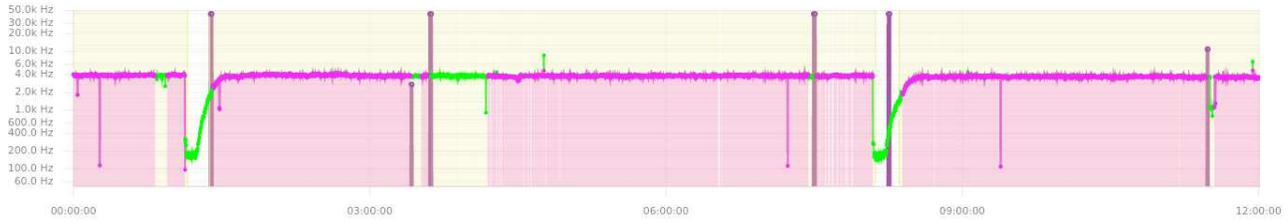}}}
  \caption{A typical half-day profile of the input (green, open histogram)
    and recorded (magenta, shaded) trigger rate, and data acquisition
    errors (vertical spikes).  The drop and recovery of the input trigger
    rate corresponds to the loss of the beam and refill, while the lack of
    the output trigger rate corresponds to dead time due to an error. }
  \label{fig:half-day}
\end{figure*}

The trigger rate is still much lower than our design, but the background
level is close to the highest level that detector can endure, as it is
the largest limiting factor of the accelerator and detector operation.
Hence the occupancy and the radiation effects to the frontend electronics are
rather severe, causing various kind of instabilities.
Fig.~\ref{fig:half-day} shows an example of the trigger rate of about
4$\,$kHz with several beam losses and recovery periods from data acquisition
errors in a half day.  Some of 
the data acquisition errors are due to immaturity of the firmware which has been
diligently improved as the commissioning went on, while some are due to
unstable hardware modules or connections which were replaced or fixed
when it was possible.

\subsection{Readout Latency}

We measure the latency of the data processing inside FEE as a part of
the unified readout system, by including the timestamp of the start of
the data transfer into the data header itself.  This can be then compared with
the timestamp of the trigger in the data stream in an offline analysis.

% We measure the latency of the data processing inside FEE as a part of
% the unified readout system, by including the timestamp of sending the
% data header into the data header itself.  This can be then compared with
% the timestamp of the trigger in the data stream in an offline analysis.

\begin{figure}[ht]
  \centerline{%
    \resizebox{0.49\textwidth}{!}{%
      \includegraphics{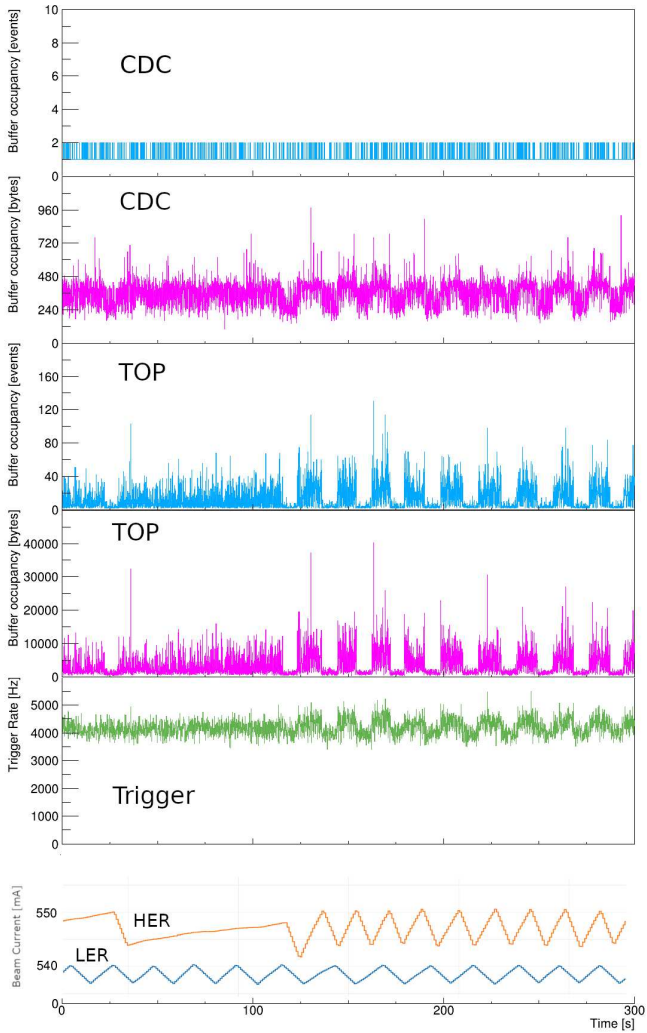}}}
  \caption{Estimated buffer occupancy in terms of number of events and
    amount of bytes per link for CDC and TOP as a function of time from
    the beginning of a run (shown for the first 5 minutes), the level-1
    trigger rate, and the beam current in the high-energy ring (HER, for
    electrons) and low-energy ring (LER, for positrons) for the same
    period.  The large occupancy period corresponds to the HER injection
    period (when the current is increased).}
  \label{fig:event-size}
\end{figure}

Fig.~\ref{fig:event-size} shows the estimated buffer occupancy at the
COPPER using this data latency, assuming the the buffer is swiftly read
out at the ideal bandwidth of the COPPER board.  The event fragment
stays inside the FIFO buffer until all data of four links are aligned.
Therefore the occupancy illustrates the typical size of event fragments
and variation of the processing time in the FEE.

We find the CDC data latency to be the smallest and almost uniform, thanks
to the single-board FEE configuration.  We also find the TOP data
latency is the largest and least uniform, as a result of software data
processing in the Arm core of the FEE.

We also use this information to extrapolate to the 30 kHz design trigger
rate to confirm that the COPPER buffer will not overflow.

\section{Problems, Errors and Solutions}

As already described, various problems and errors in the data acquisition
system have been making a large contribution to the inefficiency of data
taking.  Most of the errors are understood and improved in 2020 with
respect to the previous year, and will be further improved in coming
runs.  Here we classify the problems and errors into four categories:
single event upset (SEU), link errors, hardware failures and other
problems.

\subsection{Single Event Upset (SEU)}
\label{sec:seu}

The FEE boards of CDC, TOP and ARICH are inside the detector and are
expected to suffer from gamma rays and neutrons.  According to the
previous studies~\cite{radtest}, the most affected parts of a typical
FEE board are the optical transceiver, which is permanently damaged by a
large dose of gamma rays, and the FPGA, whose configuration or data
memory bit is flipped by SEU caused by neutrons.  The optical
transceiver we adopted survived a 10-year equivalent integrated dose of
gamma-rays in an irradiation test, while other typical transceivers we
tested stopped working at a much smaller integrated dose.

\begin{figure}[ht]
  \centerline{%
    \resizebox{0.48\textwidth}{!}{%
      \includegraphics{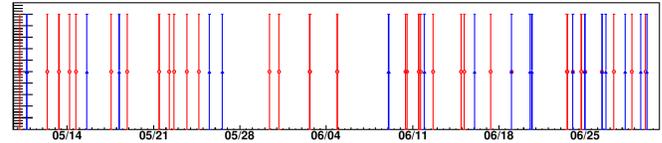}}}
  \caption{History of unrecoverable SEU errors of CDC over 50 days
    detected by the mitigation logic (blue, triangle) and detected by
    data corruption (red, circle).}
  \label{fig:seu}
\end{figure}

The CDC FEE uses the Xilinx SEU mitigation logic to correct the
configuration memory altered by SEU.  Successful SEU correction
occurs a few times a day without affecting the data acquisition, and it
is monitored through the TTD system.  However, the SEU mitigation code
is not able to correct multiple bit errors at a time or errors in the
mitigation code itself.  It does not correct the data memory either,
including those used as a part of state machines.  These unrecoverable
errors occurred at an average rate of once per day, 40\% of which were
detected by the SEU mitigation logic, as shown in
Fig.~\ref{fig:seu}.  Then the FPGA has to be reprogrammed; the
reprogramming takes less than 10$\,$s, but the detection and identification
procedure of the error currently takes a much longer time, typically
more than 5 minutes.

The TOP FEE also uses the SEU mitigation logic from Xilinx, and ARICH
uses a custom SEU mitigation logic~\cite{arich} which has a superior
performance compared with the one provided by Xilinx for the Spartan-6
FPGA.

Reduction of the down time is foreseen by automating the reprogramming
procedure of the unrecoverable SEU errors.

\subsection{Link Errors}

Both custom protocols, b2tt and Belle2link, use predefined 8b10b
control symbols to define the protocol and have an embedded data error
checking mechanism using a cyclic redundancy check (CRC).  An incorrect
control symbol and a CRC error are identified as a link error, and
propagated to the TTD system to stop the run.  The link error, either in
b2tt or Belle2link, has been so far the most frequent cause that stopped
data taking.  The error often repeated from the same link, caused by a
particular version of firmware which happened to be more timing
critical, on the line that has a smaller margin.  Unstable FEE boards
and cables were replaced to avoid the weak links during the shutdown
period to make the entire system more stable.

We have made an investigation of the electric characteristics of the
signal running on the CAT7 cables from the FTSW to the FEE during the
summer shutdown period of 2020.  We identified two particular cases that
were improved during summer, one for KLM and the other for CDC.

For KLM, we find a large sine-wave noise of around 300 kHz on the
category-7 cables.  This turned out to be due to the lack of a
proper ground connection at the FEE, and the 20$\,$m long category-7
cables between FTSW and FEE.  We have moved the location of the FTSW
modules by introducing new small VMEbus crates and reducing the cable
length to 10$\,$m, and installed a proper grounding connection at the
FEE.  Some of the LVDS drivers of the KLM FEE were damaged and were replaced
during the run in 2020; the improper ground connection is suspected to
have induced a large current from an external noise causing the damage.

\begin{figure}[ht]
  \centerline{%
    \resizebox{0.47\textwidth}{!}{%
      \includegraphics{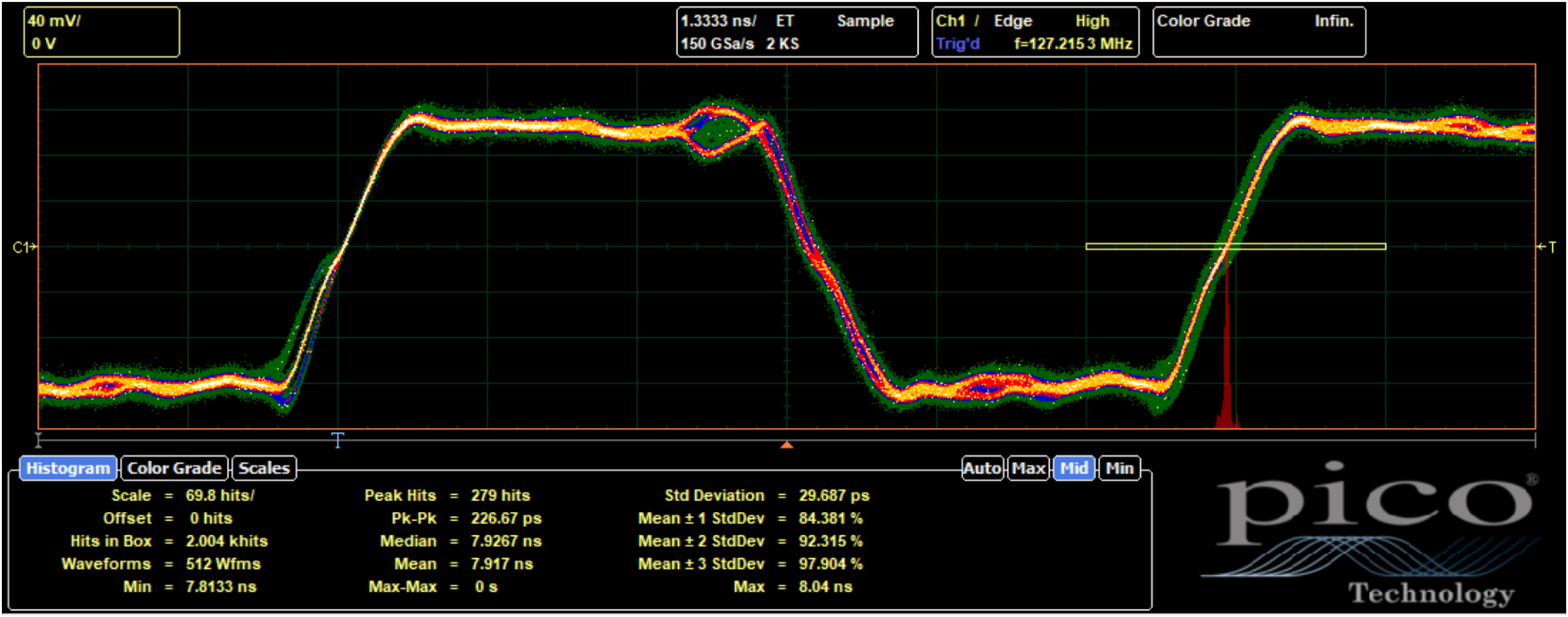}}}
  \vspace{1mm}
  \centerline{%
    \resizebox{0.47\textwidth}{!}{%
      \includegraphics{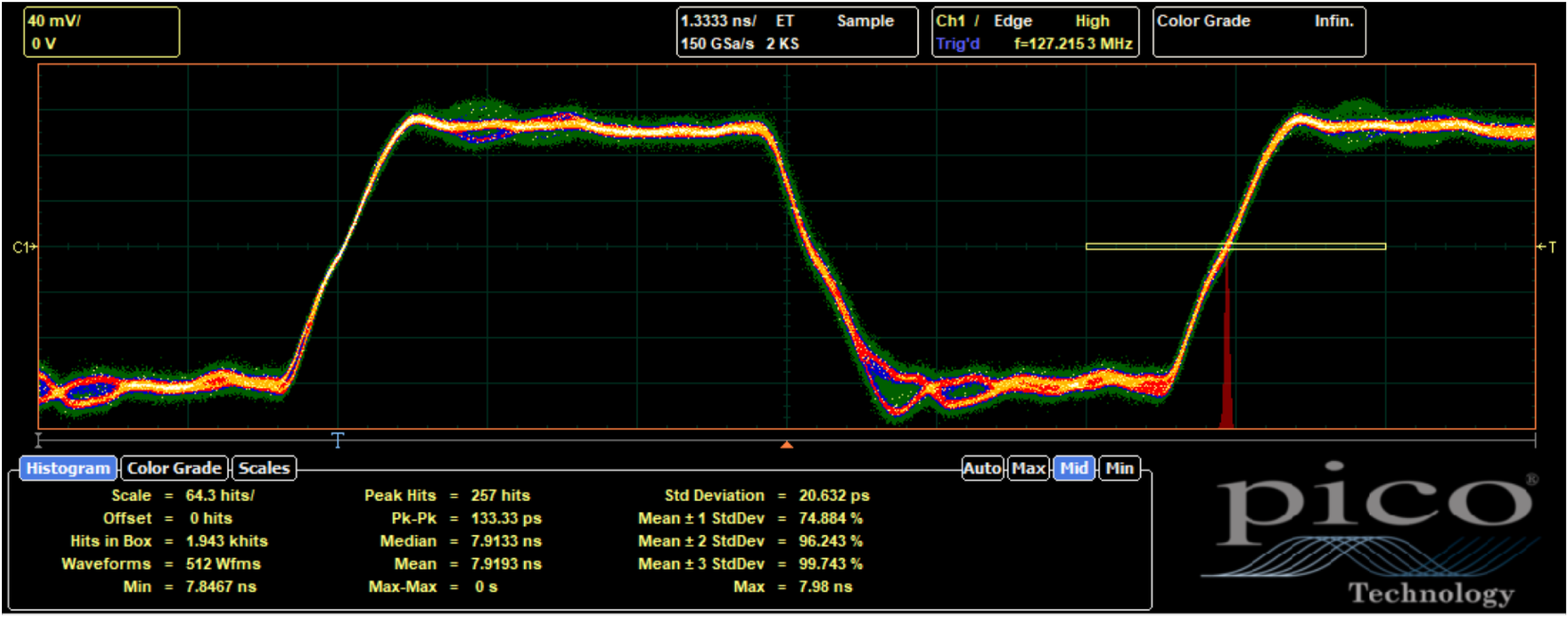}}}
  \caption{Problematic clock signal with the edge affected by the
    crosstalk (top), and improved clock 
    signal edge after adding a delay to the b2tt serial signal (bottom).}
  \label{crosstalk}
\end{figure}

The CDC FEE boards were not accessible during the summer shutdown of
2020, but the clock and serial signals were examined near the FTSW
module.  We found a combination of a lower FTSW driver amplitude and a
higher current draw at the FEE in particular connections makes it less
immune to the crosstalk from the serial-b2tt-encoded line to the clock
line.  The crosstalk causes a glitch in the clock and causes a data
error in Belle2link, although the glitch is not large enough to cause an
error in the phase lock loop (PLL).  Mostly due to this problem, up to
10 out of 299 FEE boards were masked in the worst case.  This problem
was completely cured by adding a delay to the serial link output of the
FPGA firmware, to avoid the
edge transition of the serial b2tt data near the clock edge as
shown in Fig.~\ref{crosstalk}.
% The delay is added inside the IODELAY
% function of the Virtex-5 FPGA and hence this solution was accomplished
% only by updating the FTSW firmware.

\subsection{Hardware Failures, Other Problems, and Prospects}

The largest down time occurred during the run was due to the failure in
one of the KLM FEE boards.  This FEE was a data concentrator and could
not be masked without losing a large fraction of data.  The module had
to be replaced by stopping the beam and accessing the detector area.
Other hardware failures of the FEE boards for TOP and ARICH occurred
inside the detector and it will not be possible to replace them until the long
shutdown period in 2022.

There was also down time due to the COPPER backend system, the HLT
system, and slow control software problems.

The down time of the data acquisition system is one of the major concerns
of the future run period of Belle II.  We have improved the
stability of the system in various ways at every major shutdown period
and also during the run period.  For the unavoidable errors such as the
single event upset of the FEE, we are improving the monitor and error
recovery procedure~\cite{kunigo}.

\section{Conclusions}

We have presented the performance of the unified readout system of the
Belle II experiment at the SuperKEKB $e^+e^-$ collider during the first
two years of the operation.  We have been smoothly running at about 4
kHz level-1 trigger rate with a readout dead-time fraction below 1\%.
The largest dead time is from the unavoidable continuous injection veto,
but a similarly large fraction of the dead time was caused by various
errors in the unified readout system as well as in the rest of the data
acquisition system.  We have described the major problems we encountered,
and solutions we found to improve the stability of the system and to
reduce the dead time.  We also confirmed using the real data that the
unified readout system can handle the design level-1 trigger rate of 30
kHz.  We expect a more stable operation with a higher luminosity and
trigger rate in the coming runs.

\end{document}